\documentclass[twocolumn,showpacs,preprintnumbers,amsmath,amssymb,aps,prd,nofootinbib,superscriptaddress,eqsecnum]{revtex4}
\usepackage[usenames]{color}
\usepackage{graphicx}
\usepackage{mathrsfs}
\newcommand*{\citenamefont}[1]{#1}
\newcommand*{\bibnamefont}[1]{#1}
\newcommand*{\bibfnamefont}[1]{#1}

\renewcommand\d{\partial}
\newcommand\tr{\mathop{\mathrm{Tr}}}

\newcommand\g{g_5}

\begin{document}

\title{%
Parity doubling structure of nucleon at non-zero density \\
in the holographic mean field theory
}

\author{Bing-Ran He\footnote{e-mail: he@hken.phys.nagoya-u.ac.jp}}
\author{Masayasu Harada\footnote{e-mail: harada@hken.phys.nagoya-u.ac.jp}}
\affiliation{
Department of Physics, Nagoya University, Nagoya 464-8602, Japan
}

\date{\today}

\begin{abstract}
We develope the holographic mean field approach in a
bottom-up holographic QCD model including baryons and scalar mesons
in addition to vector mesons and pions.
We study the effect of 
parity doubling structure of baryons at non-zero density
to
the equation of state between the chemical potential and 
the baryon number density.
We first show that we can adjust the amount of nucleon
mass coming from the chiral symmetry breaking by changing the boundary
value of the five-dimensional baryon fields.
Then, introducing the mean field for the baryon fields,
we
calculate the equation of state between the baryon number density and
its corresponding chemical potential. 
Then, comparing the predicted equation of state with the one obtained
in a Walecka type model, we extract the density dependence of the
effective nucleon mass. 
The result shows that the effective mass decreases with increasing
density, and that the rate of decreasing is more rapid for larger
percentage of the mass from 
the chiral symmetry breaking.
\end{abstract}


\maketitle

\section{Introduction}
\label{sec:intro}

The spontaneous chiral symmetry breaking ($\chi$SB)
is one of the most important features in 
low-energy QCD.
This is considered to be the origin of several hadron masses, 
such as the lightest nucleon mass.
However, there is a possibility that only a part of the 
lightest nucleon mass is generated by the spontaneous $\chi$SB
and the remaining part is the chiral invariant mass.
This structure is nicely expressed in so called parity doublet models
(see, e.g. 
Refs.~\cite{Detar:1988kn,Jido:2001nt,Sasaki:2010bp,Paeng:2011hy}).

It is an interesting question to ask how much amount of the
nucleon mass is generated by the spontaneous $\chi$SB,
or to investigate the origin of nucleon mass.
Studying dense baryonic matter would give some clues to
understand the origin of our mass, since a partial restoration of 
chiral symmetry will occur at high density region.
We expect that the mass generated by the spontaneous $\chi$SB
will become small near the chiral phase transition point.

It is not so an easy task to study the dense baryonic matter from the
first principle, namely starting from QCD itself:  It may not be
good to use the perturbative analysis, and the lattice QCD is 
not applicable due to the sign problem at this moment.
Then, instead of the analysis from the first principle,
it may be useful to make an analysis based on effective models,
especially for qualitative understanding.

Holographic QCD (hQCD) models (see, for reviews, e.g. 
Refs.~\cite{Erdmenger:2007cm,Kim:2012ey} and references therein.)
are constructed based on the 
AdS/CFT 
correspondence~\cite{Maldacena:1997re,Gubser:1998bc,Witten:1998qj} 
and powerful tools to study 
the low-energy hadron physics.
There exist several ways to apply hQCD models
for dense baryonic matter
(see e.g. Refs.~\cite{Kim:2006gp,Horigome:2006xu,Kobayashi:2006sb,%
Kim:2007xi,Bergman:2007wp,Seo:2008qc,Nakamura:2006xk,Rozali:2007rx}).
Recently the holographic mean field theory approach was
proposed to study dense baryonic matter in Ref.~\cite{Harada:2011aa}.
This approach allows us to predict the equation of state 
between the chemical
potential and the baryon number density.
In Ref.~\cite{Harada:2011aa}, this approach was applied to a
top-down model of hQCD~\cite{Hong:2007ay} 
including the baryon fields
in the framework of the Sakai-Sugimoto (SS) model~\cite{Sakai:2004cn}.
It is known~\cite{Dymarsky:2010ci}
that the SS model provides the
repulsive force mediated by iso-singlet mesons such as $\omega$
meson among nucleons, while the attractive force mediated by the
scalar mesons are not generated.
As a result Ref.~\cite{Harada:2011aa} shows that
the chemical potential increases monotonically with 
the baryon number density.
On the other hand, when the attraction mediated by the scalar 
meson is appropriately included, the chemical potential is 
expect to decrease up until the normal nuclear matter density,
and then turn to increase (see e.g. Ref.~\cite{Dymarsky:2010ci}).
Thus, it is interesting to study whether the chemical potential
decreases with increasing density when the scalar degree of freedom is
included. 

In this paper, for studying this, we adopt a bottom-up model given in 
Ref.~\cite{Hong:2006ta} which includes five-dimensional baryon 
field included in the model proposed in 
Refs.~\cite{Erlich:2005qh,DaRold:2005zs}.
There the five dimensional scalar field $X$ is explicitly 
included to express the chiral symmetry breaking by its vacuum
expectation value (VEV).

Yet another interest appears in a hQCD model of
Ref.~\cite{Hong:2006ta}.
Since there is no chirality in five dimension, the hQCD model includes
two baryon fields; one transforms linearly under U(2)$_{\rm R}$ and
another under U(2)$_{\rm L}$.
The existence of two baryon fields naturally 
generates the parity doublet structure mentioned above.
In Ref.~\cite{Hong:2006ta}, the boundary condition is adopted in such
a way that all of the nucleon mass is generated by the chiral symmetry
breaking.

In the present analysis, 
we will show that we can adjust the amount of nucleon
mass coming from the chiral symmetry breaking by changing the boundary
value of the five-dimensional baryon fields:
The percentages of the
chiral invariant mass in the nucleon mass
is controlled by changing the boundary value.
We study how the equation of state in the dense baryonic matter
depends on the percentage of the nucleon mass originated from the
spontaneous $\chi$SB in the holographic mean field theory approach.
Our result shows that, larger the percentage of the mass
coming from the spontaneous $\chi$SB is, more rapidly
the 
effective nucleon mass, which is extracted from the equation of state
by comparing it with the one obtained in a Walecka type model given in
Ref.~\cite{Matsui:1981ag},
with increasing baryon number density.

This paper is organized as follows:
In section~\ref{sec:parity}, we first review the model proposed
in Ref.~\cite{Hong:2006ta}, and then show the parity doubling
structure.
We study the equation of state at non-zero baryon density in the
model in section~\ref{holographic_mft}.
We also discuss the interpretation of our results in terms of a
Walecka-type model.
Finally, we give a summary and discussions in section~\ref{sec:SD}.
We summarize several intricate formulas needed in this paper
in appendix~\ref{Parity}.

\section{Parity doubling structure of the model}
\label{sec:parity}

\subsection{model}

In this subsection we briefly review the holographic QCD
model including baryons given in Ref.~\cite{Hong:2006ta}.

The fields relevant to the present analysis are
the scalar meson field $X$ and two baryon fields $N_1$ and $N_2$,
as well as the 5-dimensional gauge fields $R_A$ and
$L_A$,
which transform under the 5-dimensional chiral symmetry as
\begin{eqnarray}
X &\rightarrow& g_L \, X \,g_R^\dag \ ,\\
N_1 &\rightarrow& g_R \, N_1 \ ,\\
N_2 &\rightarrow& g_L \, N_2 \ ,\\
R_A &\rightarrow& g_R\, R_A \,g_R\dag 
  - i \partial_A g_R \cdot g_R^\dag \ , \\
L_A &\rightarrow& g_L\, L_A \,g_L^\dag 
  - i \partial_A g_L \cdot g_L^\dag \ ,
\end{eqnarray}
where $g_{R,L} \in \mbox{U}(2)_{R,L}$ denote the transformation
matrix of chiral symmetry, and $A = \mu\,,\,z$ with
$\mu= 0,\,1,\,2,\,3$.
By using these fields,
the bulk action is given as
\begin{eqnarray}
S &=& S_{N_1} + S_{N_2} + S_{\rm int} + S_X
\ ,
\label{Lagrangian}
\end{eqnarray}
where
\begin{widetext}
\begin{eqnarray}
S_{N_1}&=&\int {\rm d}^5x\sqrt{g}\, \left\{ \frac{i}{2}\bar
N_1e_A^M\Gamma^A\nabla_MN_1-\frac{i}{2}\left(\nabla_M^{\dagger}\bar
N_1\right)e_A^M\Gamma^AN_1 -M_5\bar N_1N_1 \right\} \,, 
\label{action_n1}\\
S_{N_2}&=&\int {\rm d}^5x\sqrt{g}\, \left\{ \frac{i}{2}\bar
N_2e_A^M\Gamma^A\nabla_MN_2-\frac{i}{2}\left(\nabla_M^{\dagger}\bar
N_2\right)e_A^M\Gamma^AN_2 +M_5\bar N_2N_2 \right\}\,,
\label{action_n2}\\
S_{\rm int}&=&-\int {\rm d}^5x\sqrt{g} 
G\, \left\{ \bar{N}_2XN_1 +\bar{N}_1X^\dag N_2 \right\}\,, 
\label{action_int}\\
S_X &=& = 
\int\!d^5x\,\sqrt{g}\, \tr\Big\{ |DX|^2  - {m_5}^2 |X|^2 
- \frac1{4\g^2} (F_L^2 + F_R^2) \Big\}\,,
\label{S X}
\end{eqnarray}
\end{widetext}
with $M_5=5/2$ and $m_5^2=-3$ being the bulk masses for baryons
and mesons, $G$ the scalar-baryon coupling constant,
$g_5$ the gauge coupling constant.
The vielbein $e^A_M$ appearing  in Eqs.~(\ref{action_n1}) and  
(\ref{action_n2}) satisfies  
\begin{equation}
g_{MN}=e_M^Ae_N^B\,\eta_{AB}=\frac{1}{z^2} {\rm diag}(+----)\,,
\end{equation}  
where $M$ labels the general space-time coordinate and $A$ labels the
local Lorentz space-time, with $A,M\in(0,1,2,3,z)$. 
By fixing the
gauge for the Lorentz transformation, we take the vielbein as  
\begin{equation}
e_M^A=\frac{1}{z}\eta_M^A=\frac{1}{z} {\rm diag}(+----)\,.
\end{equation}
The Dirac matrices $\Gamma^A$ 
are defined as $\Gamma^\mu=\gamma^\mu$  
and $\Gamma^z=-i\gamma^5$ which satisfy the anti-commutation relation 
\begin{equation}
\left\{\Gamma^A,\Gamma^B\right\}=2\eta^{AB}\,.
\end{equation}
The covariant derivatives for baryon and scalar meson 
are defined as
\begin{eqnarray}
\nabla_MN_1&=&(\partial_M+\frac{i}{4}\omega_M^{AB}\Gamma_{AB}-i
(A_L^a)_M t^a)N_1\,,\\ 
\nabla_MN_2&=&(\partial_M+\frac{i}{4}\omega_M^{AB}\Gamma_{AB}-i
(A_R^a)_M t^a)N_2\,,\\ 
D_M X &=&\d_M X - iA_{LM} X + iX A_{RM}\,,
\end{eqnarray}
where $\Gamma^{AB}=[\Gamma^A,\Gamma^B]/(2i)$. 
$\omega_M^{AB}$ is the spin connection 
given by
\begin{eqnarray}
\omega_M^{AB}&=&
\frac1z (\eta^A_{\;\;Z}\eta^B_{\;\;M}-\eta^A_{\;\;M}\eta^B_{\;\;Z})
\eta^{ZZ}\,.
\end{eqnarray}

\subsection{parity doubling structure}

In this subsection,
we study the parity doubling structure of baryons in the
model described in the previous subsection.
Note that the analysis in this subsection is done for 
zero chemical potential, so that only the scalar field $X$ 
has a mean field part, or 
4-dimensional vacuum expectation value (VEV), expressed by $X_0$.
The equation of motion (EoM) for $X_0$ is read from
the action $S_X$ in Eq.~(\ref{S X}) as 
\begin{equation}
\frac{1}{z^3} \d_z^2 X_0 -  \frac{3}{z^4} \d_z X_0 +  
 \frac{3}{z^5} X_0 = 0 \,.
\label{x_vev_eq}
\end{equation}
The solution for this EoM is obtained 
as~\cite{Erlich:2005qh,DaRold:2005zs}
\begin{equation}
  X_0(z) = \frac{1}{2}M
       z + \frac{1}{2} \sigma  z^3, \label{eq:vev}
\end{equation}
where $M$ is the {current quark mass} and $\sigma$ is the 
quark condensate $\langle \bar{q} q \rangle$.
By using this vacuum solution, 
the EoMs for $N_1$ and $N_2$ are given by
\begin{eqnarray}
\left(ie_A^M\Gamma^A\nabla_M-M_5\right)N_1 - GX_0N_2=0\,,\\
\left(ie_A^M\Gamma^A\nabla_M+M_5\right)N_2 - GX_0N_1=0\,.
\end{eqnarray}

As done in Ref.~\cite{Hong:2006ta}, we 
decompose the bulk fields $N_1$ and $N_2$ as
\begin{eqnarray}
N_1&=&N_{1L}+N_{1R}\,,\nonumber\\
N_2&=&N_{2L}+N_{2R}\,,\label{5dseparate2}
\end{eqnarray}
where 
\begin{align}
N_{1L}=i\Gamma^zN_{1L}\;&,\;N_{1R}=-i\Gamma^zN_{1R}\,,\nonumber\\
N_{2L}=i\Gamma^zN_{2L}\;&,\;N_{2R}=-i\Gamma^zN_{2R}\,.
\end{align}
The mode expansions of $N_{1L,R}$ and $N_{2L,R}$ are performed as
\begin{eqnarray}
N_{1L,R}(x,z)=\sum_n\int\frac{d^4p}{(2\pi)^4}e^{-ipx}f_{1L,R}^{(n)}(z)
\psi_{L,R}^{(n)}(p)
\,,\nonumber\\ 
N_{2L,R}(x,z)=\sum_n\int\frac{d^4p}{(2\pi)^4}e^{-ipx}f_{2L,R}^{(n)}(z) 
\psi_{L,R}^{(n)}(p)\, .
\nonumber\\
\label{KKmode}
\end{eqnarray}
It is convenient to introduce
$f_+^{(n)}$ and $f_-^{(n)}$ as
\begin{eqnarray}
f_{+1}^{(n)}&=&f_{1L}^{(n)} + f_{2R}^{(n)} 
\,,\nonumber\\
f_{+2}^{(n)}&=&f_{1R}^{(n)} - f_{2L}^{(n)} 
\,,\nonumber\\
f_{-1}^{(n)}&=&f_{1L}^{(n)} - f_{2R}^{(n)} 
\,,\nonumber\\
f_{-2}^{(n)}&=&f_{1R}^{(n)} + f_{2L}^{(n)} 
\,,\label{pmmode_12}
\end{eqnarray}
which satisfy
\begin{eqnarray}
\d_z f_{+1}^{(n)}& =& \frac{2+M_5}{z} f_{+1}^{(n)} 
  - \frac12 G\sigma z^2 f_{+2}^{(n)} - m_{+}^{(n)} f_{+2}^{(n)}
\,,\nonumber\\
\d_z f_{+2}^{(n)}& =&\frac{2-M_5}{z} f_{+2}^{(n)} 
  - \frac12 G\sigma z^2 f_{+1}^{(n)} + m_{+}^{(n)} f_{+1}^{(n)}
\,,
\nonumber\\
\label{eomf+}
\end{eqnarray}
and
\begin{eqnarray}
\d_z f_{-1}^{(n)}& = &\frac{2+M_5}{z} f_{-1}^{(n)}
{+} \frac12 G\sigma z^2 f_{-2}^{(n)} - m_{-}^{(n)} f_{-2}^{(n)}\,,\nonumber\\
\d_z f_{-2}^{(n)}& = &\frac{2-M_5}{z} f_{-2}^{(n)} 
{+} \frac12 G\sigma z^2 f_{-1}^{(n)} + m_{-}^{(n)} f_{-1}^{(n)}
\,,
\nonumber\\
\label{eomf-}
\end{eqnarray}
with $m_{\pm}^{(n)}$ corresponding to mass eigenvalues.

It should be noticed that
Eq.~(\ref{eomf-}) is rewritten as
\begin{eqnarray}
\partial_z f_{-1}^{(n)} & = \frac{2+M_5}{z} f_{-1}^{(n)} 
 {-} \frac12 G\sigma z^2 f_{-2}^{(n)} - (-m_{-}^{(n)}) (-f_{-2}^{(n)})
\,,\nonumber\\
\d_z (-f_{-2}^{(n)}) & = \frac{2-M_5}{z} (-f_{-2}^{(n)} )
  {-} \frac12 G\sigma z^2 f_{-1}^{(n)} + (-m_{-}^{(n)}) f_{-1}^{(n)}
\,,\nonumber\\
\label{eomf--}
\end{eqnarray}
which is the same form as in Eq.~(\ref{eomf+}).
This implies that the solutions of Eq.~(\ref{eomf+}) and 
those of Eq.~(\ref{eomf-}) are not independent with each other.
For example, a solution of Eq.~(\ref{eomf+}) with negative
energy eigenvalue is actually a solution of Eq.~(\ref{eomf-})
with positive energy eigenvalue,
which is the reflection of the 
charge conjugation invariance at zero density.

For solving Eq.~(\ref{eomf+}) we need to fix the boundary conditions
for $f_{+1}^{(n)}$ and $f_{+2}^{(n)}$:
At the UV boundary ($z=0$), $f_{+1}^{(n)}$ and $f_{+2}^{(n)}$ 
should be zero required by the normalizability.
The value of $f_{+1}$ at the IR boundary can be set $1$ without loss
of generality since the coupled differential equations 
in Eq.~(\ref{eomf+}) are homogeneous equations.
In Ref.~\cite{Hong:2006ta}, the value of $f_{+2}$ at the IR boundary
was taken as $0$ in such a way 
that all of the mass of ground state baryon is generated by 
the chiral symmetry breaking expressed by the VEV of $X_0$.

In the present analysis,
we regard the IR value of $f_{+2}$, 
i.e. $f_{+2}|_{z=z_m} = c_1$,
as a parameter, which turns out to control
the percentages of the
chiral invariant mass included in the nucleon mass.
We summarize the boundary condition in Table~\ref{tab:1}
for a convenience.
\begin{table}[htb]\vspace{-6pt}
\centering
\begin{tabular}{c|cc}  \hline\hline
        & UV    & IR\\\hline
$f_{+1}$ &0 	& $1$ \\
$f_{+2}$ &0	& $c_1$\\
\hline\hline
\end{tabular}
\caption{Boundary conditions for baryon fields}
\label{tab:1}
\end{table}
In the remaining part of this subsection, 
we shall show the dependence of the percentage of 
the chiral invariant mass of the nucleon
on the IR boundary value $c_1$, for fixed value of 
$z_m = 1/0.3236\,\mbox{(GeV)}^{-1}$~\cite{Erlich:2005qh}.

\begin{figure}[htb]
\begin{center}
\includegraphics[scale=0.42]{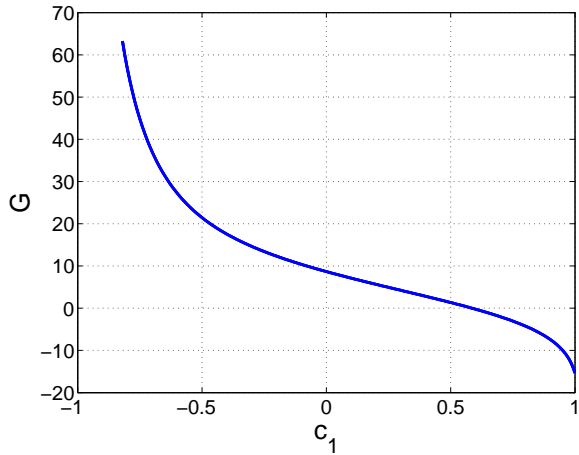}
\caption{Value of $G$ determined from $c_1$ 
to make the lowest eigenvalue to be 
the nucleon mass of $0.94\,\mbox{GeV}$.
\\
}
\label{figB_G1}
\end{center}
\end{figure}

For a given value of $c_1$, we first adjust the coupling $G$ 
to ensure that the lowest eigenvalue becomes the nucleon mass
of $0.94\,\mbox{GeV}$.
We show how the value of $G$ changes depending on the value
of $c_1$ in Fig.~\ref{figB_G1}.
It should be noted that the lowest eigenvalue is obtained by shooting $0.94\,\mbox{GeV}$ through Eq.~(\ref{eomf+}) or equivalently by shooting $-0.94\,\mbox{GeV}$ through  Eq.~(\ref{eomf-}).
We can show that, the eigenvalues obtained from Eq.~(\ref{eomf+}) with positive sign are the masses
of the excited nucleons with positive parity, while
the ones with negative sign are masses of 
negative-parity excited nucleons.
For Eq.~(\ref{eomf-}), the parity assignment is
interchanged.

We give an intricate discussion of the parity assignment in 
appendix~\ref{Parity}.

\begin{figure}[htb]
\begin{center}
\includegraphics[scale=0.42]{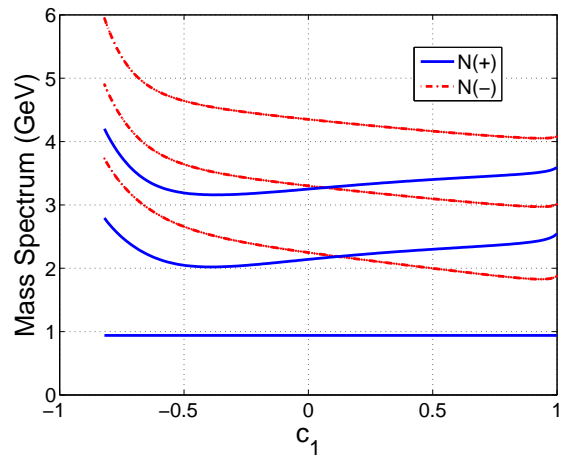}
\caption{$c_1$ dependence of excited nucleon masses.
\\
}
\label{figG_boundaryMS}
\end{center}
\end{figure}

We next calculate the masses of higher excited nucleons using
the value of $G$ determined above for fixed $c_1$.
We show the $c_1$-dependence of several masses in 
Fig.~\ref{figG_boundaryMS}.
Here, $N(+)$ denotes the states with positive parity
while $N(-)$ stands for negative parity.
This figure shows that, for $c_1 > c_1^\ast\approx 0.12$,
 the first excited state carries the negative parity and the second 
the positive parity, and so on.
For $c_1 < c_1^\ast$, on the other hand,
the first excited state is the positive-parity excited nucleon,
which seems consistent with the experimental data.

\begin{figure}[htb]
\begin{center}
\includegraphics[scale=0.42]{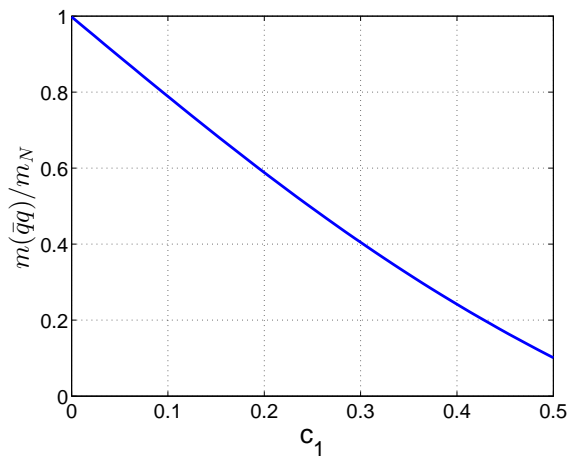}
\caption{$c_1$-dependence of 
$\displaystyle\frac{m(\bar{q}q)}{m_N}$, which shows
the percentage
of the nucleon mass coming from the spontaneous chiral symmetry 
breaking.
\\
}
\label{figBoundary_p0}
\end{center}
\end{figure}

Finally in this subsection,
we investigate the effect of dynamical chiral symmetry breaking 
on the nucleon mass.
For quantifying this effect, 
we take $\sigma=0$ and calculate the mass eigenvalue by solving
\begin{eqnarray}\label{eomf+normal_nochiral}
\partial_z f_{+1}^{(n)}
  & =& \frac{2+M_5}{z} f_{+1}^{(n)}  - {m}_0^{(n)} f_{+2}^{(n)}
\,,\nonumber\\
\partial_z f_{+2}^{(n)}
  & =& \frac{2-M_5}{z} f_{+2}^{(n)}  + {m}_0^{(n)} f_{+1}^{(n)}\,,
\end{eqnarray}
for several choices of $c_1$.
We consider the lowest eigenvalue ${m}_0^{(1)}$, denoted as
just $m_0$, as the chiral 
invariant mass of nucleon.
In Fig.~\ref{figBoundary_p0}, we plot the $c_1$ dependence
of the value of $1- m_0/m_N \equiv \frac{m(\bar{q}q)}{m_N} $ 
which shows the percentage
of the nucleon mass coming from the spontaneous chiral symmetry 
breaking.
{}From Fig.~\ref{figBoundary_p0} we conclude that,
in the case of $c_1=0$, which is chosen in Ref.~\cite{Hong:2006ta},
all the nucleon mass  comes from the spontaneous
chiral symmetry breaking. 
On the other hand, when 
$c_1 > 0.25$,
more than half
of the nucleon mass is the chiral invariant mass.

\section{Equation of state in the holographic mean field approach to
  the model }
\label{holographic_mft}

In this section, we study the finite density system using
the holographic mean field theory proposed 
in Ref.~\cite{Harada:2011aa}.
In the holographic mean field theory,
all the 5D fields are decomposed into the mean fields which depend
only on the 5th coordinate $z$ and the fluctuation fields.
In the present analysis, 
we consider the symmetric nuclear matter, so
that the proton and the neutron have the same mean fields.
Furthermore, 
we assume that the mean fields for
the vector and axial-vector gauge fields except the U(1)$_{\rm V}$
gauge field and the traceless part of the scalar field are zero.
Then, in the mean field analysis,
we make the following replacements:
\begin{eqnarray}
X(x,z)&\rightarrow&X(z)\,\mathbb{I}\,,\nonumber\\
V_M(x,z)&\rightarrow&V_\mu(z)\,\frac{1}{2}\,\mathbb{I} \,,\nonumber\\
A_M(x,z)&\rightarrow&0 \,,\nonumber\\
N_1(x,z)&\rightarrow&N_1(z) 
 \,
  \left(\begin{array}{c}
    1 \\ 1 \\
  \end{array}\right)
\,,\nonumber\\
N_2(x,z)&\rightarrow&N_2(z)\,
 \,  \left(\begin{array}{c}
    1 \\ 1 \\
  \end{array}\right)\,,
\label{HMF}
\end{eqnarray}
where $\mathbb{I}$ is the $2\times2$ unit matrix in the flavor space,
and 
  the proton and the neutron
have the same mean fields consistently with the symmetric nuclear
matter. 
Note that both $N_1(z)$  and $N_2(z)$ are four-component spinors
in the above expression.
\begin{equation}
N_1(z) = \left(\begin{array}{c}
 N_{11}(z) \\ N_{12}(z) \\ N_{13}(z) \\ N_{14}(z) \\
\end{array}\right)
\ , \quad
N_2(z) = \left(\begin{array}{c}
 N_{21}(z) \\ N_{22}(z) \\ N_{23}(z) \\ N_{24}(z) \\
\end{array}\right)
\ .
\end{equation}
The equations of motion for the mean fields,
$X(z)$, $V_M(z)$, $N_1(z)$, $N_2(z)$ are 
given by
\begin{eqnarray}
z^2\d_z^2 X -3z\d_z X - m_5^2 X -\frac G2(\bar{N}_2 N_1 + \bar{N}_1 N_2)&=&0\,,\nonumber\\
\eta^{\mu\nu}(z^3 \d_z^2 V_\nu - z^2\d_z V_\nu) - g_5^2 (\bar{N}_1 \Gamma^\mu N_1 + \bar{N}_2 \Gamma^\mu N_2)&=&0\,,\nonumber\\
\left(zi\Gamma^z\d_z-2i\Gamma^z+z\Gamma^\mu V_\mu -M_5\right)N_1 - GXN_2&=&0\,,\nonumber\\
\left(zi\Gamma^z\d_z-2i\Gamma^z+z\Gamma^\mu V_\mu +M_5\right)N_2 -
GXN_1&=&0\,.
\nonumber\\
\label{5dmeanfield} 
\end{eqnarray}

One can easily show that $V_1=V_2=0$ together with
$N_{11}=N_{13}=N_{21}=N_{23}=0$ provides a solution for the above
equations of motion.  
Furthermore,
$V_3 = 0$ becomes a solution when the
baryonic mean fields satisfy either of the following conditions: 
\begin{eqnarray}
N_{12}&=&{}-N_{24}\,,\nonumber\\
N_{14}&=&N_{22}\,,\label{N+12}
\end{eqnarray}
or 
\begin{eqnarray}
N_{12}&=&N_{24}\,,\nonumber\\
N_{14}&=&{}-N_{22}\,.\label{N-12}
\end{eqnarray}
Thus, in the following, we study the solution with
$V_1=V_2=V_3=0$ and either of Eq.~(\ref{N+12}) or
Eq.~(\ref{N-12}).

Now, it is convenient to introduce 
\begin{eqnarray}
N_{+}&=&N_{12} + N_{14}\,,\nonumber\\
N_{-}&=&N_{12} - N_{14}\,.\label{N+}
\end{eqnarray}
Then, by using Eq.~(\ref{N+12}) with $V_1=V_2=V_3=0$,
the equations of motion are rewritten as
\begin{eqnarray}
\partial_z^2 X &=&\frac3z \d_z X + \frac{m_5^2}{z^2} X + 
  \frac{G}{2z^2}(N_{+}^\dag N_{+} - N_{-}^\dag N_{-})\,,\nonumber\\
\d_z^2 V_0 &=&\frac1z \d_z V_0 + \frac{g_5^2}{z^3}(N_{+}^\dag N_{+} + N_{-}^\dag N_{-})\,,\nonumber\\
\d_z N_{+} &=& \frac{2 + M_5}{z} N_{+} - \frac1z GX N_{-} - V_0 N_{-}\,,\nonumber\\
\d_z N_{-} &=& \frac{2 - M_5}{z} N_{-} - \frac1z GX N_{+} + V_0
N_{+}\,.\label{eomn+} 
\end{eqnarray}
The equations of motions corresponding to Eq.~(\ref{N-12})
are obtained by changing the sign in front of $G$.
The situation is similar to the one for Eqs.~(\ref{eomf+})
and (\ref{eomf-}): The solutions of Eq.~(\ref{eomn+})
is connected to the one for Eq.~(\ref{eomf+}), 
and the one corresponding to Eq.~(\ref{N-12}) is to the one
for Eq.~(\ref{eomf-}).

Let us consider the boundary condition to solve the equations of
motion in Eq.~(\ref{eomn+}).
First of all, 
$V_0$ at UV boundary corresponds to the chemical potential $\mu$:
\begin{equation}
V_0(z=0)=\mu\,.
\label{def:mu}
\end{equation}
The derivative at IR boundary is taken to be zero:
$\partial_z V_0(z) \vert_{z=z_m}= 0$.
In the present analysis we do not include the effect of
current quark mass, so that the value of $X$ at UV boundary
is taken to be zero: $X(z=0)=0$.
There is an ambiguity for the IR value of $X$.
In this analysis, following Ref.~\cite{Kim:2007xi},
we fix it to be the value determined at vacuum:
$X(z=z_m)= \sigma_0 z_m^3/2$ with $\sigma_0 = (318\,\mbox{MeV})^3$.
This is based on the assumption that the IR values are not affected
much by the chemical potential introduced at the UV boundary.
For the baryon fields, we take the UV values of $N_+$ and $N_-$
to be zero following the holographic mean field 
theory~\cite{Harada:2011aa}.
The equations of motion for mean fields are not homogeneous
equations, so that the normalization of the baryon fields become
relevant.
We change the IR values of $N_+$ and $N_-$ to control the
baryon number density,
which is written in terms of the baryon fields as~\footnote{
We checked that the baryon number density defined as in Eq. (3.9)
agrees with the one defined from the UV value of the gauge field as 
$\partial_z V_0 |_{z=0}$.
}
\begin{equation}
\rho_{\rm b}  = \int \frac{dz}{2z^4} (N_+^\dag N_+ + N_-^\dag N_-) = \int dz \,\rho(z)
\, .
\label{numberdensity_b}
\end{equation}
It should be noted that the ratio of two baryon fields at IR
boundary is left free as in the previous section.
So, we use
$N_+(z=z_m)=c_2$ and $N_-(z=z_m)=c_2 \times c_1$,
where $c_2$ determines the baryon number density while
$c_1$ controls the percentage of the chiral invariant mass
of nucleon.
We summarize the boundary conditions in Table~\ref{muspectboundary}.
\begin{table}[htb]
\centering
\begin{tabular}{c|cc}  \hline\hline
       & UV    & IR\\\hline
$X$    & $0$   & $\sigma_0 z_m^3/2$ \\
$V_0$  & $\mu$ & - \\
$\partial_z V_0$ & - & $0$ \\
$N_{1}$ & $0$   & $c_2$\\
$N_{2}$ & $0$   & $c_2*c_1$\\\hline\hline
\end{tabular}
\caption{Boundary condition at finite density.
The mark ``{-}'' indicates that the value is not fixed.}
\label{muspectboundary}
\end{table}

We solve the equations of motion in Eq.~(\ref{eomn+})
for given values of $c_1$ and $c_2$,
with regarding $\mu$ in Eq.~(\ref{def:mu}) as an eigenvalue.
Using the solutions for the baryonic mean fields $N_+$ and $N_-$
we calculate the baryon number density from
Eq.~(\ref{numberdensity_b}).

\begin{figure}[htb]
\begin{center}
\includegraphics[scale=0.42]{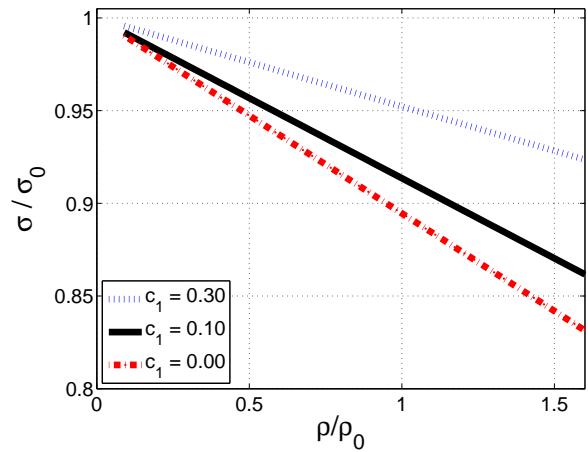} 
\caption{Density dependence of $\sigma/\sigma_0$ for several choices
  of $c_1$.
}
\label{figND_sigma}
\end{center}
\end{figure}

We first study the density dependence of the chiral condensate for checking the partial chiral restoration.  Here we
define the in-medium condensate
through the holographic mean field $X(z)$ as
\begin{equation}\label{sigma_z}
\left.\sigma=\frac{2X(z)}{z^3} \right\vert_{z=z_{\rm UV}}\,.
\end{equation}
We plot the density dependence of the $\sigma$ normalized by 
the vacuum value $\sigma_0$ in Fig.~\ref{figND_sigma}.
This shows that the quark condensate $\sigma$
decreases with the increasing number density,
which can be regarded as a sign of the 
partial chiral symmetry restoration.
When the value of $c_1$ is decreased,
the corresponding
value of $G$ becomes larger (see Fig.~\ref{figB_G1})
to reproduce the nucleon mass.  
Since the larger $G$ implies the larger correction to the scalar
from the nucleon matter,
the smaller $c_1$ we choose, the more rapidly the condensate $\sigma$
decreases.  
The degreasing property of the chiral condensate 
is similar to the one 
obtained in Ref.~\cite{Kim:2007xi}.

\begin{figure}[htb]
\begin{center}
\includegraphics[scale=0.42]{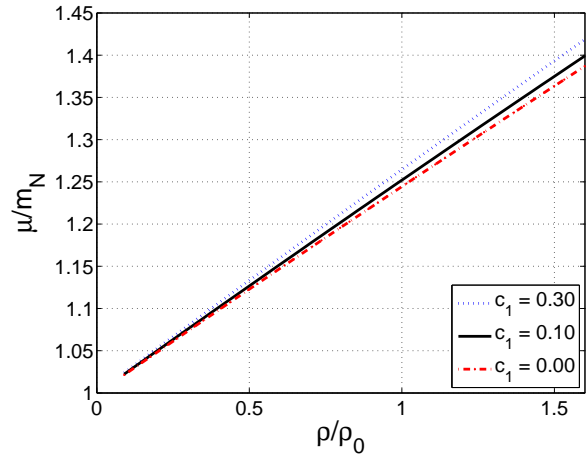} 
\caption{Equation of state. The horizontal axis shows the baryon
  number density normalized by the normal nuclear matter density
of $\rho_0=0.16\,{\rm (fm)}^{-3}$, 
and the vertical axis does the chemical potential by the 
nucleon mass of $0.94$\,GeV.
The dashed line shows the EoS for $c_1=0$, the solid line for
$c_1=0.1$ and the dotted line for $c_1=0.3$.
}
\label{figND_mu}
\end{center}
\end{figure}

We next show the resultant equation of state, a relation between 
the chemical potential and the baryon number density in 
Fig.~\ref{figND_mu}.
This figure shows that the chemical potential increases with the
increasing baryon number density.
This does not agree with the nature, in which the chemical potential decreases against the density in the low density region below the normal nuclear matter density.
This decreasing property is achieved by the subtle cancellation between the repulsive and attractive forces.
So this increasing property indicates that, in the present model, the repulsive force mediated by the U(1) gauge field is stronger than the attractive force mediated by the scalar degree included in X field.

\begin{figure}[htb]
\begin{center}
\includegraphics[scale=0.42]{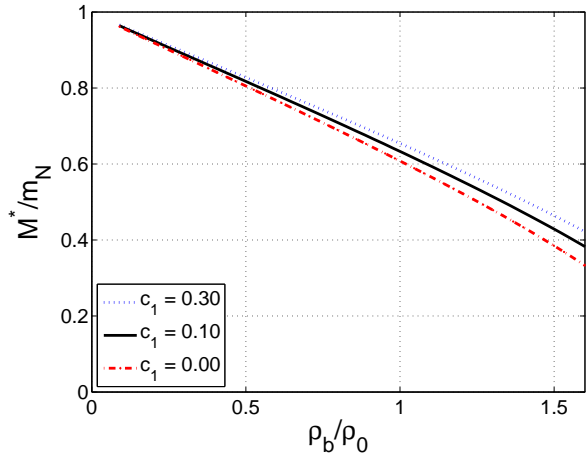}
\caption{
Density dependence of the effective nucleon mass $M^\ast$.
}
\label{fig : effective mass}
\end{center}
\end{figure}

For studying the attractive force mediated by the scalar fields, we extract the density dependence of the effective nucleon mass using the Walecka type model
(see e.g. Refs.~\cite{Walecka:1974qa,Matsui:1981ag}),
in which the chemical potential $\mu$ is expressed as
\begin{equation}
\mu = \sum_{n=1}^{\infty}\frac{g_{\omega^{(n)} NN}^2}{m_{\omega^{(n)}}^2} \rho_b 
  + \sqrt{ k_F^2 + {M^{\ast}}^2 }
\ ,
\label{walecka}
\end{equation}
where $\rho_b$ is the baryon number density, $g_{\omega^{(n)} NN}$ is the coupling for $n$th eigenstate of the omega mesons, $m_{\omega^{(n)}}$ is its mass, $k_F$ is the
Fermi momentum, and $M^\ast$ is the effective nucleon mass.
Note that, in the free Fermi gas, $k_F$ is related to $\rho_b$ as
$
\rho_b = \frac{2 k_F^3}{3 \pi^2}
$,
which leads to
$
k_F = \left( \frac{3 \pi^2 \rho_b}{2} \right)^{1/3}
$.
In the present hQCD model, 
the $\omega^{(n)} NN$ coupling is calculated in vacuum as
$g_{\omega^{(n)} NN}= 15.5\sim15.8,\,8.9\sim10.9\ldots$ depending on the value of $c_1$.
Using these couplings
together with the masses of $m_{\omega^{(n)}} \sim 780,\,1794\,\ldots\mbox{MeV}$,
we convert the density dependence of $\mu$ obtained 
above into the one of the
effective nucleon mass $M^\ast$ through Eq.~(\ref{walecka}).
We plot the density dependence of the effective mass $M^\ast$ in 
Fig.~\ref{fig : effective mass}.
This shows that the effective mass decreases with increasing density.
The decreasing rate is larger than the one obtained in 
Ref.~\cite{Kim:2007xi}, which is the reflection of the
iterative corrections included through the holographic
mean field theory.
It should be noted that the decreasing of $M^\ast$ is more
rapid for smaller value of $c_1$.  In other word, the larger the
percentage of the mass coming from the chiral symmetry breaking is,
more rapidly the effective mass $M^\ast$ decreases with density.

\begin{figure}[htb]
\begin{center}
\includegraphics[scale=0.42]{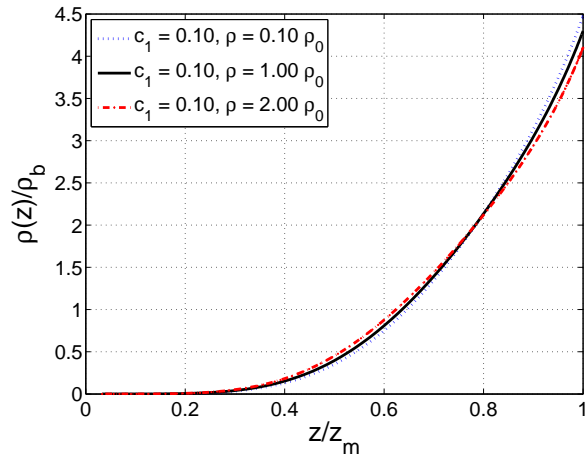} 
\caption{
Baryon charge distribution
$\rho(z)/\rho_b$.
}
\label{fig : distribution}
\end{center}
\end{figure}

In Fig.~\ref{fig : distribution}, we plot the baryon charge 
distribution $\rho(z)$ defined in Eq.~(\ref{numberdensity_b})
for $\rho=0.1\,\rho_0$, $\rho_0$ and $2\,\rho_0$ with $c_1 = 0.1$ fixed.
This figure shows that the distribution is broader for larger value of $\rho$.  This indicates that the distribution becomes more important for larger density, as shown in Ref.~\cite{Harada:2011aa}.

\section{A summary and discussions}
\label{sec:SD}

We develope the holographic mean field approach in a
bottom-up holographic QCD model proposed in Ref.~\cite{Hong:2006ta}
which includes five-dimensional baryon  
field in the model proposed in 
Refs.~\cite{Erlich:2005qh,DaRold:2005zs}.
We first study the mass spectrum of baryons with paying attention to
the chiral invariant mass $m_0$, which were formulated in parity
doublet models
(see, e.g. 
Refs.~\cite{Detar:1988kn,Jido:2001nt,Sasaki:2010bp,Paeng:2011hy}).
We found the parameter ($c_1$), which is one boundary value of two baryon fields, controls the percentage of the chiral invariant mass:
for $c_1=0$ all of the mass of the ground-state nucleon
is generated by the spontaneous chiral symmetry breaking, while for 
$ c_1  >0.25$, 
more than half of the nucleon
mass is actually the chiral invariant mass.

We studied the density dependence of the chiral condensate.
Our result shows that the quark condensate $\sigma$
decreases with the increasing number density, 
which is consistent with the analysis done in Ref.~\cite{Kim:2007xi}. 
Furthermore, we found that the $\sigma$ decreases more rapidly for
smaller value of $c_1$.
This is because the sigma coupling to the nucleon is larger for
smaller $c_1$. 

We next calculated the equation of state between the baryon
chemical potential and the baryon number density using the holographic
mean field approach proposed in Ref.~\cite{Harada:2011aa}.
The resultant equation of state shows that 
the chemical potential increases with the increasing baryon number density.
This
indicates that, in the present model, the repulsive force mediated by the U(1) gauge field is stronger than the attractive force mediated by the scalar degree included in X field.
For studying the attractive force mediated by the scalar fields, we extract the density dependence of the effective nucleon mass using a Walecka type model.
Our result shows 
that the effective mass decreases with increasing density.
Furthermore, the decreasing rate is more rapid 
for smaller 
value of $c_1$.
This is consistent with the fact that the percentage of
the chiral invariant mass is larger for larger value of 
$ c_1$. 
In other word, the
larger the percentage of the mass
coming from the spontaneous $\chi$SB is, more rapidly the effective nucleon mass decreases with increasing baryon number density.

We also studied the baryon number distribution in the holographic
direction.
Our results show that the distribution is concentrated near the
IR boundary for smaller $\rho$.
This indicates that the distribution becomes more important for larger density.

In the present analysis, we made an analysis only at 
the mean field level. 
So a natural extension is to consider 
the fluctuations on the top of the
mean field obtained here.
It is also interesting to study the relation between the isospin 
chemical potential and the isospin density based on the approach
developed in this paper, since the 
relation has a relevance to the symmetry energy.   
We leave these works to the future project.

\section*{Acknowledgement}

M.H. would like to thank useful discussions with Youngman Kim
and Chang-Hwan Lee.
We are also grateful to Yong-Liang Ma for helpful discussions.
This work 
was supported in part by Grant-in-Aid for Scientific Research
on Innovative Areas (No. 2104) ``Quest on New
Hadrons with Variety of Flavors'' from MEXT,
and by the JSPS Grant-in-Aid for Scientific Research
(S) No. 22224003, (c) No. 24540266.
BR.H. would like to thank 
the Nagoya University Program for Leading 
Graduate Schools ``Leadership Development Program for Space
Exploration and Research'' for the financial support.

\begin{appendix}

\section{Parity transformation}
\label{Parity}

In this appendix, 
we consider the parity transformation properties of the 5D fields. 
As in the 4-dimension, 
a parity transformation should flip the
sign of normal three spatial coordinates.  
But the 5th coordinate $z$, as it is 
defined in the range $z_{UV}< z <z_{IR}$, 
does not participate in the parity transformation. 
\begin{eqnarray}
x^\mu &\xrightarrow{\rm P}& x_\mu\,,\quad z ~~\xrightarrow{\rm P} ~~z\,,\nonumber\\
\d_\mu &\xrightarrow{\rm P}& \d^\mu\,,\quad \d_z ~~\xrightarrow{\rm P} ~~\d_z\,.
\end{eqnarray}
Using these conventions, we obtain the parity transformation of 5D
fields as  
\begin{eqnarray}
X(x,z)&\xrightarrow{\rm P}& X^\dagger(x,z)\,,\nonumber\\
A_{L\mu}(x,z) &\xrightarrow{\rm P}& A_{R}^{~~\mu}(x,z)\,,\nonumber\\
A_{R\mu}(x,z) &\xrightarrow{\rm P}& A_{L}^{~~\mu}(x,z)\,,\nonumber\\
A_{Lz}(x,z) &\xrightarrow{\rm P}& A_{Rz}(x,z)\,,\nonumber\\
A_{Rz}(x,z) &\xrightarrow{\rm P}& A_{Lz}(x,z)\,.\label{5dxparity}
\end{eqnarray}
For the 5D spinors, their parity transformation properties are
express as
\begin{eqnarray} 
N_{1L}(x,z) &\xrightarrow{\rm P}& \eta_1 \gamma^0 N_{2R}(x,z)
\,,\nonumber\\
N_{1R}(x,z) &\xrightarrow{\rm P}& \eta_2 \gamma^0 N_{2L}(x,z)
\,,\nonumber\\
N_{2L}(x,z) &\xrightarrow{\rm P}& \eta_2 \gamma^0 N_{1R}(x,z)
\,,\nonumber\\
N_{2R}(x,z) &\xrightarrow{\rm P}& \eta_1  \gamma^0 N_{1L}(x,z)
\,,\label{5dnparity}
\end{eqnarray}
where $\eta_1$ and $\eta_2$ are arbitrary phases.

For explicitly illustrating the parity invariance we rewrite the 5D
Lagrangian in Eqs.~(\ref{action_n1})-(\ref{action_int}) 
in terms of chiral basis as 
\begin{widetext}
\begin{eqnarray}
\mathscr{L}_{N_1}&=& \bar{N}_1\left(iz\Gamma^M\d_M -2i\Gamma^z
+z\Gamma^M A^a_{LM}t^a-M_5\right)N_1
\nonumber\\
&=&{}\bar{N}_{1L}\left(iz\Gamma^\mu\d_\mu + z\Gamma^\mu A^a_{L\mu}t^a
\right){N}_{1L} 
+\bar{N}_{1R}\left(iz\Gamma^\mu\d_\mu +z\Gamma^\mu A^a_{L\mu}t^a
\right){N}_{1R}
\nonumber\\
&&{}+\bar{N}_{1L}\left( iz\Gamma^z\d_z +z\Gamma^z A^a_{Lz}t^a
-2i\Gamma^z -M_5\right){N}_{1R}
+\bar{N}_{1R}\left( iz\Gamma^z\d_z +z\Gamma^z A^a_{Lz}t^a
-2i\Gamma^z -M_5\right){N}_{1L}
\,,\label{n1lagrangian}\\
\mathscr{L}_{N_2}&=& \bar{N}_2\left(iz\Gamma^M\d_M -2i\Gamma^z
+z\Gamma^M A^a_{RM}t^a + M_5\right)N_2
\nonumber\\
&=&\bar{N}_{2L}\left(iz\Gamma^\mu\d_\mu + z\Gamma^\mu A^a_{R\mu}t^a
\right){N}_{2L} 
+\bar{N}_{2R}\left(iz\Gamma^\mu\d_\mu +z\Gamma^\mu A^a_{R\mu}t^a
\right){N}_{2R}
\nonumber\\ 
&&{}+\bar{N}_{2L}\left( iz\Gamma^z\d_z +z\Gamma^z A^a_{Rz}t^a
-2i\Gamma^z + M_5\right){N}_{2R} 
+\bar{N}_{2R}\left( iz\Gamma^z\d_z +z\Gamma^z A^a_{Rz}t^a
-2i\Gamma^z + M_5\right){N}_{2L}
\,,\label{n2lagrangian}\\
\mathscr{L}_{\rm int}&=&{}-G[\bar{N}_2X{N}_1 + \bar{N}_1X^\dag{N}_2]
\nonumber\\
&=&{}-G[\bar{N}_{2L}X{N}_{1R} + \bar{N}_{2R}X{N}_{1L} +
  \bar{N}_{1L}X^\dag{N}_{2R} +\bar{N}_{1R}X^\dag{N}_{2L}] 
\,.\label{intlagrangian}
\end{eqnarray}
Under parity transformation, they transform as
\begin{eqnarray}
\mathscr{L}_{N_1}&\xrightarrow{\rm P}& {\eta_1}^* \eta_1 \bar{N}_{2R}\left(iz\Gamma^\mu\d_\mu + z\Gamma^\mu A^a_{R\mu}t^a \right){N}_{2R} \nonumber\\
&&+ {\eta_2}^* \eta_2 \bar{N}_{2L}\left(iz\Gamma^\mu \d_\mu +z\Gamma^\mu A^a_{R\mu}t^a \right){N}_{2L}\nonumber\\
&&- {\eta_1}^* {\eta_2} \bar{N}_{2R}\left( iz\Gamma^z\d_z +z\Gamma^z A^a_{Rz}t^a  -2i\Gamma^z + M_5\right){N}_{2L} \nonumber\\
&&- {\eta_2}^* \eta_1 \bar{N}_{2L}\left( iz\Gamma^z\d_z +z\Gamma^z A^a_{Rz}t^a   -2i\Gamma^z + M_5\right){N}_{2R}  \,,\label{n1plagrangian}\\
\mathscr{L}_{N_2}&\xrightarrow{\rm P}&  {\eta_2}^* \eta_2 \bar{N}_{1R}\left(iz\Gamma^\mu\d_\mu + z\Gamma^\mu A^a_{L\mu}t^a \right){N}_{1R} \nonumber\\
&&+{\eta_1}^* {\eta_1} \bar{N}_{1L}\left(iz\Gamma^\mu \d_\mu +z\Gamma^\mu A^a_{L\mu}t^a \right){N}_{1L}\nonumber\\
&&-  {\eta_2}^* \eta_1 \bar{N}_{1R}\left( iz\Gamma^z\d_z +z\Gamma^z A^a_{Lz}t^a  -2i\Gamma^z -M_5\right){N}_{1L} \nonumber\\
&&-  {\eta_1}^* \eta_2 \bar{N}_{1L}\left( iz\Gamma^z\d_z +z\Gamma^z A^a_{Lz}t^a   -2i\Gamma^z -M_5\right){N}_{1R}  \,,\label{n2plagrangian}\\
\mathscr{L}_{\rm int}&\xrightarrow{\rm P}& -G  [{\eta_2}^* \eta_2 (\bar{N}_{1R}X^\dag {N}_{2L} +\bar{N}_{2L}X{N}_{1R})\nonumber\\
&& \quad+ {\eta_1}^* \eta_1 (\bar{N}_{1L}X^\dag{N}_{2R} + \bar{N}_{2R}X{N}_{1L})]  \,.\label{intplagrangian}
\end{eqnarray}
It is easy to see that these actions are parity invariant
if $\eta_1$ and $\eta_2$ satisfy 
\begin{equation}
{\eta_1}^* {\eta_1}={\eta_2}^* {\eta_2} =1\;,\;
{\eta_1}^* {\eta_2} ={\eta_2}^* {\eta_1}=-1
\ .
\end{equation}  
\end{widetext}

\end{appendix}

\end{document}